\documentclass{jpsj2}

\title{
Nonequilibrium Green's-Function Approach to the Suppression of Rectification at Metal--Mott-Insulator Interfaces
}

\author{
Kenji \textsc{Yonemitsu}$^{1,2}$
\thanks{E-mail: kxy@ims.ac.jp}
}

\inst{
$^{1}$Institute for Molecular Science, Okazaki, Aichi 444-8585 \\
$^{2}$Department of Functional Molecular Science, Graduate University for Advanced Studies, Okazaki, Aichi 444-8585 
}

\abst{
The suppression of rectification at metal--Mott-insulator interfaces, which was previously shown by numerical solutions to the time-dependent Schr\"odinger equation and experiments on real devices, is reinvestigated theoretically using nonequilibrium Green's functions. The one-dimensional Hubbard model is used for a Mott insulator. The effects of attached metallic electrodes are incorporated into the self-energy. A scalar potential originating from work-function differences and satisfying the Poisson equation is added to the model. For electron density, we decompose it into three parts. One is obtained by integrating the local density of states over energy to the midpoint of the electrodes' chemical potentials. The others, obtained by integrating lesser Green's functions, are due to couplings with the electrodes and correspond to an inflow and an outflow of electrons. In Mott insulators, incoming electrons and holes are extended over the whole system, avoiding further accumulation of charges relative to that in the case without bias. This induces collective charge transport and results in the suppression of rectification. 
}

\kword{
metal-insulator interface, rectification, Poisson equation, nonequilibrium Green's function, Mott insulator
}

\begin{document}
\maketitle

\section{Introduction} 

Correlated electron systems can be candidate materials for novel functions of electronic devices. 
Most theories for electronic devices are, however, based on a conventional one-electron picture for band semiconductors or band insulators. For correlated electron systems, such a picture may need to be replaced. Devices are always surrounded by interfaces, where the match or mismatch of work functions or bands \cite{takahashi_apl06} must be considered in an appropriate manner. Generally, where two materials with different work functions are attached, bands are bent. A Schottky barrier is formed at a metal-insulator interface, \cite{schottky_38,mott_38} which is governed by the long-range Coulomb interaction. Because the barrier height is modulated by external bias, the current is usually not an odd function of external bias. The application of forward (reverse) voltage lowers (raises) the interfacial barrier, leading to a larger (smaller) current, i.e., rectifying action at the metal--band-insulator interface. \cite{sze-ng_book07} 

The importance of electron correlation in charge transport through metal-insulator interfaces is recognized in metal-insulator-semiconductor field-effect transistor (MISFET) device structures based on organic single crystals of the quasi-one-dimensional Mott insulator (BEDT-TTF)(F$_2$TCNQ) [BEDT-TTF=bis(ethylenedithio)tetrathiafulvalene, F$_2$TCNQ=2,5-difluorotetracyanoquinodimethane]. \cite{hasegawa_prb04} In Mott insulators, the carrier injections are ambipolar even if the work function of the crystal is quite different from that of the electrodes. Such characteristics are reproduced in the one-dimensional Hubbard model attached to a tight-binding model, where the formation of Schottky barriers is taken into account by added potentials satisfying the Poisson equation. \cite{yonemitsu_jpsj05} A connection was suggested between these ambipolar field-effect characteristics and the suppression of rectification at metal--Mott-insulator interfaces. \cite{yonemitsu_pacifichem} 

We have shown that rectification at metal--Mott-insulator interfaces is indeed suppressed, compared with rectification at metal--band-insulator interfaces, even for large work-function differences. \cite{yonemitsu_prb07b} This fact is demonstrated by numerical solutions to the time-dependent Schr\"odinger equation and also by experiments on real devices made of organic crystals of (BEDT-TTF)(F$_2$TCNQ). With the current-voltage characteristics $ I $=$ f(V) $ of the metal-insulator interface, it is shown that the drain current $ I_\mathrm{D} $ of the field-effect transistor is approximately proportional to $ V_\mathrm{D} f'(U_\mathrm{G}) $ for small $ V_\mathrm{D} $; $ V_\mathrm{D} $ denotes the drain voltage, $ U_\mathrm{G} $ the gate voltage in the symmetric-gate operation, \cite{hasegawa_prb04} and $ f'(V) $ is the derivative of $ f(V) $. Thus, the ambipolar field-effect characteristics [$ V_\mathrm{D} f'(U_\mathrm{G}) \sim $ an even function of $ U_\mathrm{G} $] are the consequence of the suppressed rectification [$ f(V) \sim $ an odd function of $ V $] at the metal--Mott-insulator interfaces. 

In the above-mentioned theoretical studies, after the ground state is obtained, a finite voltage is suddenly applied and maintained at a constant value. Because we employ the periodic boundary condition for a finite system (with appropriate choice of a gauge), the current under a constant bias finally oscillates owing to the finite-size effect. \cite{oka_prl03} Thus, the current density is estimated by averaging the time-dependent one over a given time period. The systems that can be treated by this method are limited to those systems in which the band structure of the left electrode coincides with that of the right electrode by shifting a constant energy. Furthermore, the time-averaged quantities are generally different from the corresponding quantities in the steady state. 

In order to avoid such artifacts, we need to treat steady states, which are free of the limitation on the metallic electrodes and the current oscillation due to the finite-seize effect. Without a heat bath, a steady state is reached by couplings with infinitely large, metallic electrodes. In studying the current-voltage characteristics in such a steady state, nonequilibrium Green's functions are common tools. Because we employ metallic electrodes where electrons are noninteracting, the effects of the electrodes can be incorporated into self-energies. \cite{meir_prl92,wingreen_prb93,jauho_prb94} They allow us to discuss the characteristics in terms of the electron density distribution. 

In this study, we employ them to reinvestigate the current-voltage characteristics caused by metal--Mott-insulator interfaces. The suppression of rectification at these interfaces can be elucidated by observing the spatial dependence of a ``nonequilibrium'' part of electron density. We will reformulate the Green's-function method in such a way that we distinguish an inflow of electrons from an outflow of electrons in the ``nonequilibrium'' part of electron density, which is crucial to the interpretation of numerical results. We will show that, in Mott insulators, incoming electrons and holes are extended over the whole system, giving rise to a collective charge transport responsible for the suppressed rectification. 

\section{Model and Method \label{sec:model_method}}

We consider an insulator, to which metallic electrodes are attached on the left and right sides.  For this central part, we use the one-dimensional Hubbard model (with on-site repulsion, $ U >$0) for a Mott insulator and the one-dimensional tight-binding model with alternating transfer integrals ($ \delta t \neq $0) for a band insulator, both at half filling: 
\begin{eqnarray}
H_{\mathrm{cen}} = \sum_{i=1}^{L_C} \left[ \psi_i n_i 
+ U (n_{i\uparrow}-1/2)(n_{i\downarrow}-1/2) \right] 
& & \nonumber \\ - \sum_{i=1}^{L_C-1} \sum_\sigma 
\left[ t_c + (-1)^i \delta t \right] \left(
c^\dagger_{i,\sigma} c_{i+1,\sigma} + c^\dagger_{i+1,\sigma} c_{i,\sigma} 
\right) 
\;, & & 
\label{eq:model}
\end{eqnarray}
where $ c^\dagger_{i,\sigma} $ ($c_{i,\sigma} $) creates (annihilates) an electron with spin $ \sigma $ at site $ i $, $ n_{i\sigma} = c^\dagger_{i,\sigma} c_{i,\sigma} $, and $ n_i = \sum_\sigma n_{i\sigma} $. $ L_C $ is the total number of sites in the central part. The scalar potential $ \psi_i $ is introduced above to account for the redistribution of electrons at interfaces to form barriers, compensating for the work-function differences $ \phi_L $ and $ \phi_R $ in equilibrium. \cite{yonemitsu_prb07b} The applied voltage $ V $ is defined such that it is positive when the right electrode has a lower potential (for electrons) than the left and the current (without multiplication of charge) flows to the right. 

Although the potential $ \psi_i $ is defined on lattice points, we solve the Poisson equation in the continuum space, 
\begin{equation}
\frac{\mathrm{d}^2 \psi}{\mathrm{d} x^2} = -V_P \left( \langle n \rangle - 1 \right) 
\;,
\label{eq:poisson}
\end{equation}
where the potential $ \psi $ and the expectation value of the electron density per site $ \langle n \rangle $ are functions of $ x $, and $ V_P $ comes from the long-range Coulomb interaction. In order to match the Fermi levels, we set the boundary condition, i.e., the potentials in the metallic electrodes, as 
\begin{eqnarray}
\psi(x) & = & 
-\phi_L + V/2 \;\;\;\mathrm{for}\; x < 1 \;, \nonumber \\ 
\psi(x) & = & 
-\phi_R - V/2 \;\;\;\mathrm{for}\; x > L_C 
\;.
\label{eq:boundary}
\end{eqnarray}
In order to solve the Poisson equation analytically, we assume 
$ -\mathrm{d}n(\psi)/ \mathrm{d}\psi = \kappa $ with a constant compressibility $ \kappa $, as in ref.~\citen{yonemitsu_prb07b} where analytic formulas are shown. There are alternative approaches tested, as explained in Appendix~\ref{sec:alternative}. We have confirmed using the expectation value $ \langle n \rangle $ obtained self-consistently that the current-voltage characteristics and the charge distributions are qualitatively unchanged. 

The effects of the left and right ($ \alpha $=$ L $, $ R $) electrodes consisting of noninteracting electrons on the central part [eq.~(\ref{eq:model})] are described generally by retarded self-energies. In the wide-band limit, they are independent of energy and their matrix elements with the site indices $ i $ and $ j $ are given by \cite{jauho_prb94}
\begin{equation}
\left( \Sigma^r_\alpha \right)_{ij} = 
-(\mathrm{i}/2) \left( \Gamma_\alpha \right)_{ij} \equiv 
-(\mathrm{i}/2) \gamma_\alpha \delta_{i,i_\alpha} \delta_{j,i_\alpha} 
\;, 
\end{equation}
where $ \delta_{ij} $=1 for $ i $=$ j $ and 0 otherwise, $ i_L $=1 denotes the site connected to the left electrode, and $ i_R $=$ L_C $ denotes the site connected to the right electrode. Within the Hartree-Fock approximation, the retarded Green's function for spin $ \sigma $, $ G^r_\sigma (\epsilon) $, is given by 
\begin{equation}
\left[ G^r_\sigma(\epsilon)^{-1} \right]_{ij} = \epsilon \delta_{i,j} - 
\left( H^r_{\mathrm{HF}\sigma} \right)_{ij} 
\;, 
\end{equation}
with 
\begin{equation}
\left( H^r_{\mathrm{HF}\sigma} \right)_{ij} = 
\left( H_{\mathrm{HF}\sigma} \right)_{i,j} 
-(\mathrm{i}/2) \sum_{\alpha=L,R} \gamma_\alpha \delta_{i,i_\alpha} \delta_{j,i_\alpha} 
\;, 
\end{equation}
where the diagonal elements of $ H_{\mathrm{HF}\sigma} $ are given by 
\begin{equation}
\left( H_{\mathrm{HF}\sigma} \right)_{i,i} = 
\psi_i + U \langle n_{i\bar{\sigma}}-1/2 \rangle
\;, 
\end{equation}
with $ \bar{\sigma} $=$ -\sigma $, and the off-diagonal elements are 
\begin{equation}
\left( H_{\mathrm{HF}\sigma} \right)_{i,i+1} = 
\left( H_{\mathrm{HF}\sigma} \right)_{i+1,i} = 
-\left[ t_c + (-1)^i \delta t \right] 
\;, 
\end{equation}
and $ \left( H_{\mathrm{HF}\sigma} \right)_{i,j} $=0 for $ \mid i-j \mid > 1 $. 

We numerically solve the eigenvalue equation \cite{datta_book95}
\begin{equation}
\sum_{j=1}^{L_C} \left( H^r_{\mathrm{HF}\sigma} \right)_{ij} 
u^\sigma_m (j) = 
\left( \epsilon^\sigma_m - \mathrm{i} \gamma^\sigma_m/2 \right) u^\sigma_m (i) 
\;, \label{eq:eigenvalue}
\end{equation}
where $ \epsilon^\sigma_m - \mathrm{i} \gamma^\sigma_m/2 $ ($\epsilon^\sigma_m $ and $ \gamma^\sigma_m $ are real) is an eigenvalue of the complex symmetric matrix $ H^r_{\mathrm{HF}\sigma} $ corresponding to the right eigenvector  $ u^\sigma_m (i)  $. We can set up the eigenvalue equation for the adjoint matrix \cite{datta_book95} \begin{equation}
\sum_{j=1}^{L_C} \left( H^a_{\mathrm{HF}\sigma} \right)_{ij} 
v^\sigma_m (j) = 
\left( \epsilon^\sigma_m + \mathrm{i} \gamma^\sigma_m/2 \right) v^\sigma_m (i) 
\;, 
\end{equation}
where $ H^a_{\mathrm{HF}\sigma} $ is the Hermitian conjugate of $ H^r_{\mathrm{HF}\sigma} $. The $ u $'s and $ v $'s are not identical because $ H^r_{\mathrm{HF}\sigma} $ and $ H^a_{\mathrm{HF}\sigma} $ are non-Hermitian matrices. Because $ H^r_{\mathrm{HF}\sigma} $ and $ H^a_{\mathrm{HF}\sigma} $ are symmetric matrices, we can take $ v^\sigma_m (i) $=$ u^{\sigma\ast}_m (i) $. In other words, the left eigenvector is given by the complex conjugate of the right one. They are functions of densities $ \langle n_{i\sigma} \rangle $ ($ i $=1, $ \cdots $, $ L_C $; $ \sigma $=$ \uparrow $, $ \downarrow $) and must be determined self-consistently. The eigenvectors are normalized according to 
\begin{equation}
\sum_i v^{\sigma\ast}_m (i) u^\sigma_n (i) = 
\sum_i u^\sigma_m (i) u^\sigma_n (i) = \delta_{mn}
\;. 
\end{equation}
For the numerical solutions presented below, we can always find a complete set of eigenvectors satisfying  
\begin{equation}
\sum_m u^\sigma_m (i) v^{\sigma\ast}_m (j) = 
\sum_m u^\sigma_m (i) u^\sigma_m (j) = \delta_{ij}
\;. 
\end{equation}
Then, the retarded Green's function is written as 
\begin{equation}
\left[ G^r_\sigma(\epsilon) \right]_{ij} = 
\sum_m \frac{ u^\sigma_m (i) u^\sigma_m (j) }
{ \epsilon - \epsilon^\sigma_m + \mathrm{i} \gamma^\sigma_m/2 }
\;. \label{eq:retarded_green}
\end{equation}

For the density $ \langle n_{i\sigma} \rangle $, we modify the frequently used formula in ref.~\citen{brandbyge_prb02} so as to maintain symmetry concerning simultaneous particle-hole transformation ($ c_{i,\sigma} \rightarrow (-1)^i c^\dagger_{i,\sigma}  $) and space-inversion operation ($ i \rightarrow L_C + 1 - i $) for $ \gamma_L $=$ \gamma_R $ and $ \phi_L $+$ \phi_R $=0 ($ L_C $ is assumed to be an even number). The modification is also useful for distinguishing between an inflow and an outflow of electrons, as will be shown later. To be more precise, we decompose the density into the ``equilibrium'' part and the parts due to the couplings with the left and right electrodes: 
\begin{equation}
\langle n_{i\sigma} \rangle = n^{\mathrm{eq}}_{i\sigma} 
+ \sum_{\alpha=L,R} \delta n^\alpha_{i\sigma} 
\;, \label{eq:density}
\end{equation} 
where the ``equilibrium'' part is defined by integrating the local density of states over energy to the midpoint of the left and right chemical potentials, $ \mu_C = ( \mu_L + \mu_R )/2 $. When the left (right) chemical potential is higher, $ \delta n^L_{i\sigma} $ ($ \delta n^R_{i\sigma} $) corresponds to the inflow, and the other corresponds to the outflow. The ``equilibrium'' part is then written as 
\begin{equation}
n^{\mathrm{eq}}_{i\sigma} \equiv \int_{-\infty}^{\infty} \mathrm{d} \epsilon
\left\{ -\frac{1}{\pi} \mathrm{Im} \left[ G^r_\sigma(\epsilon) \right]_{ii}
\right\} f_C(\epsilon) 
\;, \label{eq:density_eq}
\end{equation}
where $ f_C(\epsilon) $ is the Fermi distribution function in a virtual system with the chemical potential $ \mu_C $. Considering zero temperature, we substitute the step function for the Fermi distribution function: $ f_C (\epsilon) = \theta ( \mu_C - \epsilon ) $. The present decomposition is general and useful irrespective of whether or not the wide-band limit is applied. In general cases, the energy dependence of $ \Gamma_\alpha $ should be kept below. \cite{jauho_prb94} The definition in ref.~\citen{brandbyge_prb02} corresponds to the setting of $ \mu_C $ at either $ \mu_L $ or $ \mu_R $. 

For the ``nonequilibrium'' part of the density, remember that the lesser self-energy is given in the wide-band limit by \cite{jauho_prb94} 
\begin{equation}
\Sigma^<_\sigma (\epsilon) = \mathrm{i} \left[ 
\Gamma_L f_L (\epsilon) + \Gamma_R f_R (\epsilon) \right]
\;, 
\end{equation}
with $ f_\alpha (\epsilon) = \theta ( \mu_\alpha - \epsilon ) $. It can be decomposed again into the ``equilibrium'' part and the parts due to the couplings with the left and right electrodes: 
\begin{equation}
\Sigma^<_\sigma (\epsilon) = \Sigma^{<\mathrm{eq}}_\sigma (\epsilon) 
+ \sum_\alpha \delta \Sigma^{<\alpha}_{\sigma} (\epsilon) 
\;, 
\end{equation}
with 
\begin{equation}
\Sigma^{<\mathrm{eq}}_\sigma (\epsilon) = 
\mathrm{i} ( \Gamma_L + \Gamma_R ) f_C (\epsilon) 
\;, 
\end{equation}
and 
\begin{equation}
\delta \Sigma^{<\alpha}_{\sigma} (\epsilon) = 
\mathrm{i} \Gamma_\alpha \left[ f_\alpha (\epsilon) - f_C (\epsilon) \right] 
\;. \label{eq:delta_lesser_sigma}
\end{equation}
Because the contribution from the ``equilibrium'' part $ \Sigma^{<\mathrm{eq}}_\sigma (\epsilon) $ is regarded as included in $ n^{\mathrm{eq}}_{i\sigma} $, the ``nonequilibrium'' part of the density, $ \delta n^\alpha_{i\sigma} $, can be defined as 
\begin{equation}
\delta n^\alpha_{i\sigma} \equiv \frac{1}{2\pi \mathrm{i}} \int_{-\infty}^{\infty} 
\mathrm{d} \epsilon \left[ \delta G^{<\alpha}_{\sigma} (\epsilon) \right]_{ii}
\;, \label{eq:density_noneq}
\end{equation}
where the ``nonequilibrium'' part of the lesser Green's function, $  \delta G^{<\alpha}_{\sigma} (\epsilon) $, is given by the Keldysh equation 
\begin{equation}
\delta G^{<\alpha}_{\sigma} (\epsilon) = 
G^r_\sigma(\epsilon)
\delta \Sigma^{<\alpha}_{\sigma} (\epsilon) 
G^a_\sigma(\epsilon)
\;, 
\end{equation}
with $ \delta \Sigma^{<\alpha}_{\sigma} (\epsilon) $ in eq.~(\ref{eq:delta_lesser_sigma}) and $ G^a_\sigma(\epsilon) $ being the Hermitian conjugate of $ G^r_\sigma(\epsilon) $. 

Using eq.~(\ref{eq:retarded_green}), we will derive each part of eq.~(\ref{eq:density}). The local density of states is written as 
\begin{eqnarray}
& & -\frac{1}{\pi} \mathrm{Im} \left[ G^r_\sigma(\epsilon) \right]_{ii} 
\nonumber \\ & = & \sum_m 
\mathrm{Re} \left[ u^\sigma_m (i) \right]^2 
\frac{1}{\pi} \frac{\gamma^\sigma_m/2}
{( \epsilon - \epsilon^\sigma_m )^2+( \gamma^\sigma_m/2 )^2}
\nonumber \\ & - & \sum_m 
\mathrm{Im} \left[ u^\sigma_m (i) \right]^2 
\frac{1}{\pi} \frac{ \epsilon - \epsilon^\sigma_m }
{( \epsilon - \epsilon^\sigma_m )^2+( \gamma^\sigma_m/2 )^2} 
\;. \label{eq:dos}
\end{eqnarray}
When we substitute eq.~(\ref{eq:dos}) into eq.~(\ref{eq:density_eq}), the second term of eq.~(\ref{eq:dos}) gives a logarithmic term. This is an artifact of the energy-independent imaginary parts of the eigenvalues in eq.~(\ref{eq:eigenvalue}) (i.e., the wide-band limit), so that it is ignored below. The first term of eq.~(\ref{eq:dos}) gives 
\begin{equation}
n^{\mathrm{eq}}_{i\sigma} = \sum_m 
\mathrm{Re} \left[ u^\sigma_m (i) \right]^2 \left[ \frac{1}{\pi} \tan^{-1} 
\frac{2(\mu_C-\epsilon^\sigma_m)}{\gamma^\sigma_m} + \frac12 \right]
\;. \label{eq:formula_eq}
\end{equation}
The terms in the bracket above are reduced to the step function in the limit of $ \gamma^\sigma_m \rightarrow $0 (i.e., $ \gamma_\alpha \rightarrow $0). 

After we substitute eq.~(\ref{eq:retarded_green}) into eq.~(\ref{eq:density_noneq}), we obtain the ``nonequilibrium'' part by the integral ranges from $ \mu_C $ to $ \mu_\alpha $: 
\begin{eqnarray}
\delta n^\alpha_{i\sigma} & = & 
\frac{1}{2\pi} \int_{-\infty}^{\infty} \mathrm{d} \epsilon 
\left[ G^r_\sigma(\epsilon) \Gamma_\alpha G^a_\sigma(\epsilon) \right]_{ii} 
\left[ f_\alpha (\epsilon) - f_C (\epsilon) \right] 
\nonumber \\ & = &
\frac{\gamma_\alpha}{2\pi} \int_{-\infty}^{\infty} \mathrm{d} \epsilon 
\mid \left[ G^r_\sigma(\epsilon) \right]_{ii_\alpha}
\mid^2 \left[ f_\alpha (\epsilon) - f_C (\epsilon) \right] 
\nonumber \\ & = &
\frac{\gamma_\alpha}{2\pi} \int_{\mu_C}^{\mu_\alpha} \mathrm{d} \epsilon \sum_{n,m}
\frac{ u^\sigma_m (i) u^\sigma_m (i_\alpha) 
u^{\sigma\ast}_n (i) u^{\sigma\ast}_n (i_\alpha) }
{ \epsilon^\sigma_m - \epsilon^\sigma_n 
- \mathrm{i} \gamma^\sigma_m/2  - \mathrm{i} \gamma^\sigma_n/2 }
\nonumber \\ & \times & 
\left( \frac{1}{\epsilon - \epsilon^\sigma_m + \mathrm{i} \gamma^\sigma_m/2} - 
\frac{1}{\epsilon - \epsilon^\sigma_n - \mathrm{i} \gamma^\sigma_n/2} \right)
\nonumber \\ & = &
\frac{\gamma_\alpha}{2\pi} \sum_{n,m} 
\left\{
\mathrm{Im}\left(
\frac{ u^\sigma_m (i) u^\sigma_m (i_\alpha) 
u^{\sigma\ast}_n (i) u^{\sigma\ast}_n (i_\alpha) }
{ \epsilon^\sigma_m - \epsilon^\sigma_n 
- \mathrm{i} \gamma^\sigma_m/2  - \mathrm{i} \gamma^\sigma_n/2 } \right) \right.
\nonumber \\ & \times & 
\left[   \tan^{-1} \frac{2(\mu_\alpha - \epsilon^\sigma_m)}{ \gamma^\sigma_m } 
       - \tan^{-1} \frac{2(\mu_C - \epsilon^\sigma_m)}{ \gamma^\sigma_m } \right.
\nonumber \\ & & 
\left. + \tan^{-1} \frac{2(\mu_\alpha - \epsilon^\sigma_n)}{ \gamma^\sigma_n } 
       - \tan^{-1} \frac{2(\mu_C - \epsilon^\sigma_n)}{\gamma^\sigma_n } \right]
\nonumber \\ & + & 
\mathrm{Re}\left(
\frac{ u^\sigma_m (i) u^\sigma_m (i_\alpha) 
u^{\sigma\ast}_n (i) u^{\sigma\ast}_n (i_\alpha) }
{ \epsilon^\sigma_m - \epsilon^\sigma_n 
- \mathrm{i} \gamma^\sigma_m/2  - \mathrm{i} \gamma^\sigma_n/2 } \right)
\nonumber \\ & \times & 
\left[   \frac12 \ln \frac{ (\mu_\alpha - \epsilon^\sigma_m)^2+( \gamma^\sigma_m /2)^2 }
{ (\mu_C - \epsilon^\sigma_m)^2+( \gamma^\sigma_m /2)^2 } \right.
\nonumber \\ & & \left.
\left. - \frac12 \ln \frac{ (\mu_\alpha - \epsilon^\sigma_n)^2+( \gamma^\sigma_n /2)^2 }
{ (\mu_C - \epsilon^\sigma_n)^2+( \gamma^\sigma_n /2)^2 } \right]
\right\}
\;. \label{eq:formula_noneq}
\end{eqnarray}
This expression ensures that $ \delta n^\alpha_{i\sigma} $ is a real quantity, in contrast to the formula in ref.~\citen{brandbyge_prb02}. In the atomic limit, eqs.~(\ref{eq:formula_eq}) and (\ref{eq:formula_noneq}) are simplified, as shown in Appendix~\ref{sec:atomic_limit}.

Finally, by using the formula in ref.~\citen{haug_book08} with $ e $=$ \hbar $=1, the current from the left electrode is given by 
\begin{eqnarray}
J & = & 
\int_{-\infty}^{\infty} \frac{\mathrm{d} \epsilon}{2\pi} \sum_\sigma \mathrm{Tr}
\left[ \Gamma_L G^r_\sigma(\epsilon) \Gamma_R G^a_\sigma(\epsilon) \right] 
\left[ f_L(\epsilon) - f_R(\epsilon) \right] 
\nonumber \\ & = & 
\frac{ \gamma_L \gamma_R }{ 2\pi } \int_{-\infty}^{\infty} \mathrm{d} \epsilon 
\sum_\sigma \mid \left[ G^r_\sigma(\epsilon) \right]_{1L_C} \mid^2 
\left[ f_L(\epsilon) - f_R(\epsilon) \right] 
\nonumber \\ & = & 
\frac{ \gamma_L \gamma_R }{ 2\pi } \int_{-\infty}^{\infty} \mathrm{d} \epsilon 
\sum_\sigma 
\left\{
\mid \left[ G^r_\sigma(\epsilon) \right]_{L_C1} \mid^2 
\left[ f_L(\epsilon) - f_C(\epsilon) \right] \right. \nonumber \\ & & \left. 
- 
\mid \left[ G^r_\sigma(\epsilon) \right]_{1L_C} \mid^2 
\left[ f_R(\epsilon) - f_C(\epsilon) \right] 
\right\}
\nonumber \\ & = & 
\sum_\sigma \left( 
\gamma_R \delta n^L_{L_C\sigma} - \gamma_L \delta n^R_{1\sigma} 
\right)
\;. \label{eq:current}
\end{eqnarray}
Therefore, the current is expressed by the ``nonequilibrium'' parts of the density. We need to substitute 
$ \mu_L $=$ V/2 $, $ \mu_R $=$ -V/2 $, and $ \mu_C $=0 into all the equations above. 

\section{Results \label{sec:result}}

\begin{figure}[tb]
\includegraphics[height=12cm]{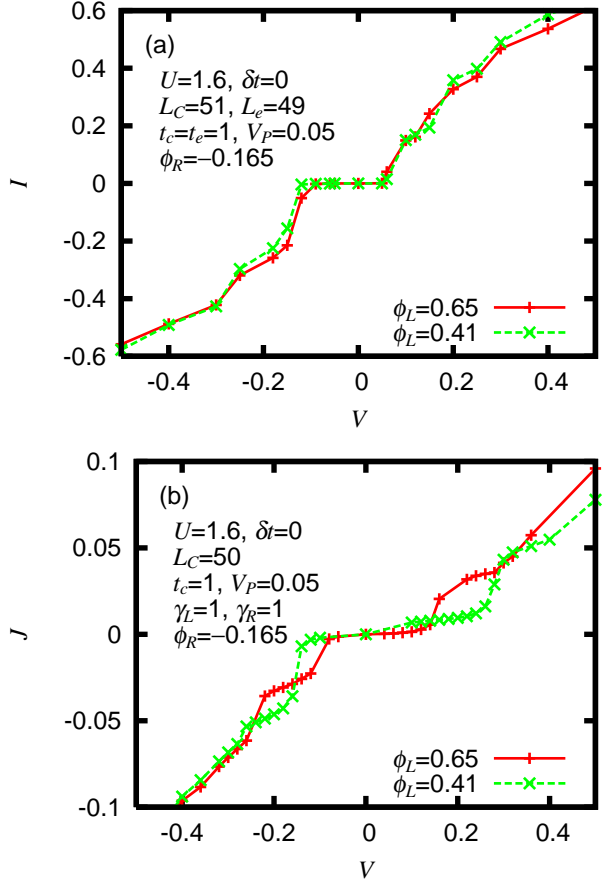}
\caption{(Color online) Current-voltage characteristics of Mott insulator with $ U $=1.6, $ \delta t $=0, $ t_c $=1, $ V_P $=0.05, and $ \phi_R $=$-$0.165 with different $ \phi_L $ values as indicated, (a) obtained by the time-dependent Hartree-Fock approximation ($ L_C $=51, $ L_e $=49, and $ t_e $=1), \cite{yonemitsu_prb07b} and (b) obtained by nonequilibrium Green's functions ($ L_C $=50 and $ \gamma_L $=$ \gamma_R $=1).
\label{fig:HFMW_IV}}
\end{figure}

Before showing the density distributions of electrons, we compare the current-voltage characteristics directly obtained by time evolution through a constant potential difference with those obtained by the nonequilibrium Green's functions in the wide-band limit. As in our previous study, \cite{yonemitsu_prb07b} we have numerically solved the time-dependent Schr\"odinger equation for the insulator sandwiched between two metallic electrodes with different work functions and a common bandwidth. Figure~\ref{fig:HFMW_IV}(a) shows the current-voltage characteristics thus obtained for the Mott insulator. 
The parameters are introduced and the current density $ I $ is defined in ref.~\citen{yonemitsu_prb07b} and $ L_C = L - L_e $. The rectifying action is shown to be suppressed. Because the bandwidths of the metallic electrodes are finite, the current density $ I $ tends to become saturated for a large $ V $. 

\begin{figure}[tb]
\includegraphics[height=12cm]{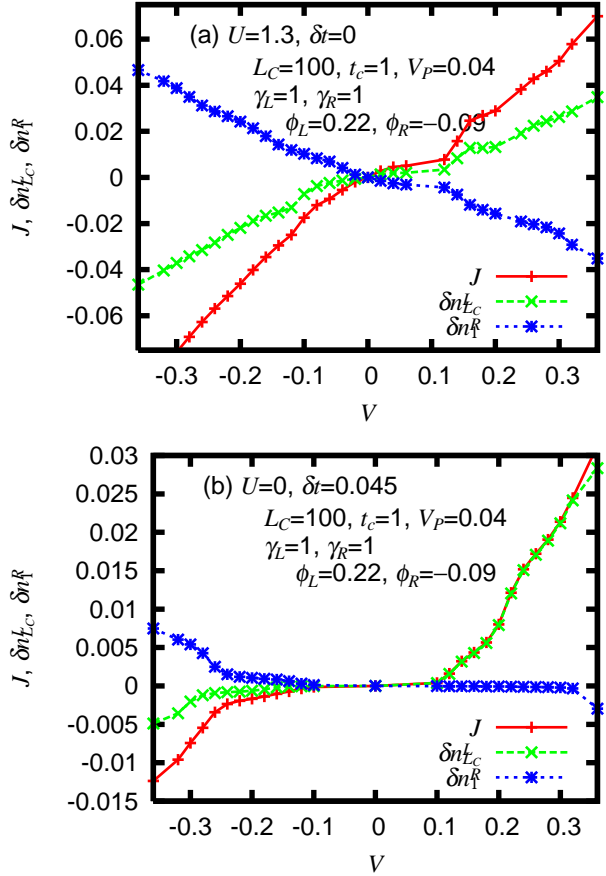}
\caption{(Color online) Current-voltage characteristics, $ \delta n^L_{L_C} $, and $ \delta n^R_1 $, of (a) Mott insulator with $ U $=1.3 and $ \delta t $=0, and (b) band insulator with $ U $=0 and $ \delta t $=0.045. The other parameters are $ L_C $=100, $ t_c $=1, $ V_P $=0.04, $ \gamma_L $=$ \gamma_R $=1, $ \phi_L $=0.22, and $ \phi_R $=$-$0.09.
\label{fig:MW_IdnV}}
\end{figure}

Figure~\ref{fig:HFMW_IV}(b) shows the current-voltage characteristics obtained by the present approach with the nonequilibrium Green's functions for the Mott insulator. The rectifying action is shown to be suppressed. Because the bandwidths of the metallic electrodes are assumed to be infinite, no saturation is observed. The values of the other parameters $ U $, $ \delta t $, $ t_c $, $ V_P $, $ \phi_L $, and $ \phi_R $ are the same as those in Fig.~\ref{fig:HFMW_IV}(a), and $ L_C $ is close to that used in Fig.~\ref{fig:HFMW_IV}(a). However, the finite metallic electrodes in Fig.~\ref{fig:HFMW_IV}(a) are replaced by infinite ones in Fig.~\ref{fig:HFMW_IV}(b), making direct comparison difficult. As a general trend, the current-voltage characteristics directly obtained by time evolution show smoother curves possibly because the current density is averaged over a given time period. As the system size $ L_C $ increases, the present approach gives smoother characteristics, as shown in Fig.~\ref{fig:MW_IdnV}. In any case, the suppression of rectification for Mott insulators is observed by both of the methods discussed above in a wide parameter space spanned by the coupling strength $ U $, the system size $ L_C $, the Coulomb parameter $ V_P $, and the work-function differences $ \phi_L $ and $ \phi_R $.

\begin{figure}[tb]
\includegraphics[height=12cm]{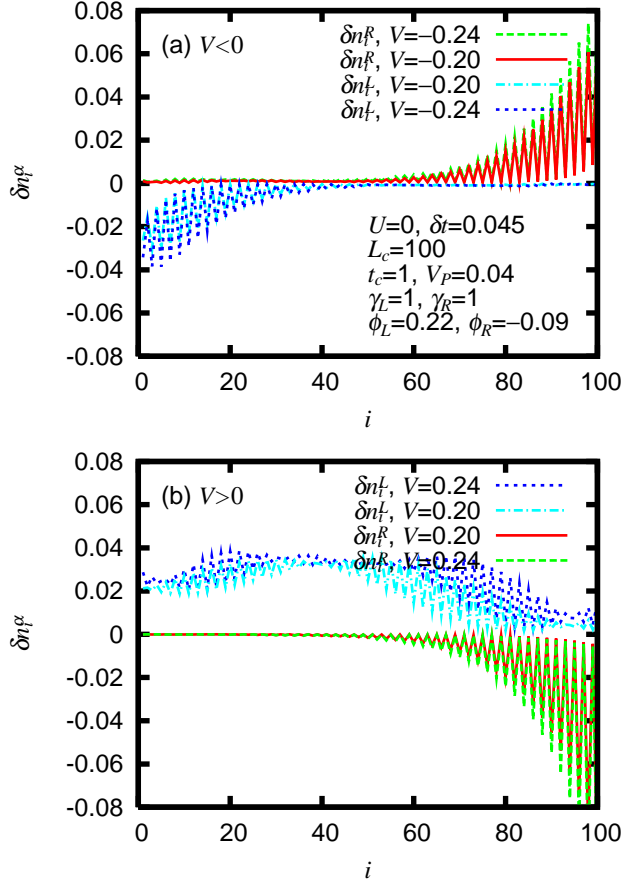}
\caption{(Color online) ``Nonequilibrium'' part of density for (a) $ V < 0$ ($ \delta n^R_i $ for $ V $=$-$0.24, $ \delta n^R_i $ for $ V $=$-$0.20, $ \delta n^L_i $ for $ V $=$-$0.20, and $ \delta n^L_i $ for $ V $=$-$0.24 from top to bottom) and (b) $ V > 0 $ ($ \delta n^L_i $ for $ V $=0.24, $ \delta n^L_i $ for $ V $=0.20, $ \delta n^R_i $ for $ V $=0.20, and $ \delta n^R_i $ for $ V $=0.24 from top to bottom) in band insulator with $ U $=0 and $ \delta t $=0.045. The other parameters are the same as those in Fig.~\ref{fig:MW_IdnV}. 
\label{fig:band_various_V}}
\end{figure}

Figure~\ref{fig:MW_IdnV} shows the current $ J $, the ``nonequilibrium'' part of the density due to the coupling with the left electrode at the rightmost site $ \delta n^L_{L_C} $, and that due to the coupling with the right electrode at the leftmost site  $ \delta n^R_1 $, which are related by $ J = \gamma_R \delta n^L_{L_C} - \gamma_L \delta n^R_1 $ in eq.~(\ref{eq:current}).
In the antisymmetric case of $ \gamma_L $=$ \gamma_R $ and $ \phi_L $=$ - \phi_R $ (not shown), the present approach guarantees the symmetry concerning simultaneous particle-hole transformation and space-inversion operation, which leads to $ \delta n^L_{L_C+1-i} $=$ - \delta n^R_{i} $ for any $ i $. In a more general case of $ \phi_L \neq - \phi_R $ but $ \gamma_L $=$ \gamma_R $, Fig.~\ref{fig:MW_IdnV} shows that the approximate relation $ \delta n^L_{L_C} \sim - \delta n^R_1 $ still holds for only Mott insulators with $ \phi_L $ and $ \phi_R $ used here. In the band insulator, the absolute value of $ J $ is larger for $ V>0 $ [Fig.~\ref{fig:MW_IdnV}(b)], where the barrier is lower than that for $ V<0 $, as expected. Later in Fig.~\ref{fig:band_various_V}, we will see how the rectification is realized in the band insulator. 

\begin{figure}[tb]
\includegraphics[height=6cm]{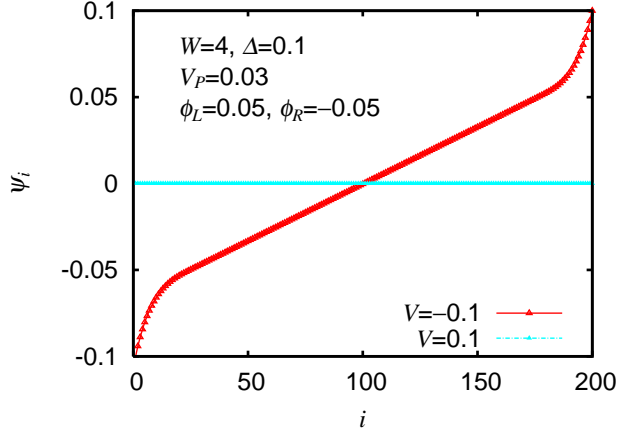}
\caption{(Color online) Scalar potential $ \psi_i $ for $ V $=$-$0.1 (with barriers for left-going bias) and $ V $=0.1 (without barriers for right-going bias), obtained analytically with the assumption of a constant compressibility. The other parameters are $ L_C $=200, the bandwidth $ W $=4, $ \Delta $=0.1, $ V_P $=0.03, and $ \phi_L $=$ -\phi_R $=0.05. 
\label{fig:potential}}
\end{figure}

From now on, unless otherwise stated, we compare Mott and band insulators with a gap $ \Delta $=0.1, which is much smaller than the values used in previous studies. \cite{yonemitsu_jpsj05,yonemitsu_pacifichem,yonemitsu_prb07b} If we numerically solved the time-dependent Schr\"odinger equation as before for such a small gap, we would need to adopt a small $ V $ to calculate the evolution for a long time, which is proportional to $ L_C/V $. Here, we use $ L_C $=200 (larger than before) and $ V_P $=0.03 (slightly smaller than before), which would also need long calculations in the previous approach. Because the present study focuses on the density distributions of incoming and outgoing electrons, we fix the parameter set to show them. Unless otherwise stated, the work function of the left (right) electrode is set to match the bottom (top) of the upper (lower) Hubbard or conduction (valence) band, $ \phi_L $=$ -\phi_R $=$ \Delta/2 $. Thus, the barriers at the two interfaces almost or completely disappear for the right-going bias ($ V >$0), and they become prominent for the left-going bias ($ V <$0), as shown in Fig.~\ref{fig:potential}. 
The absolute values of $ J $ are large for $ V >$0 and small for $ V <$0 for the band insulator. 

\begin{figure}[tb]
\includegraphics[height=12cm]{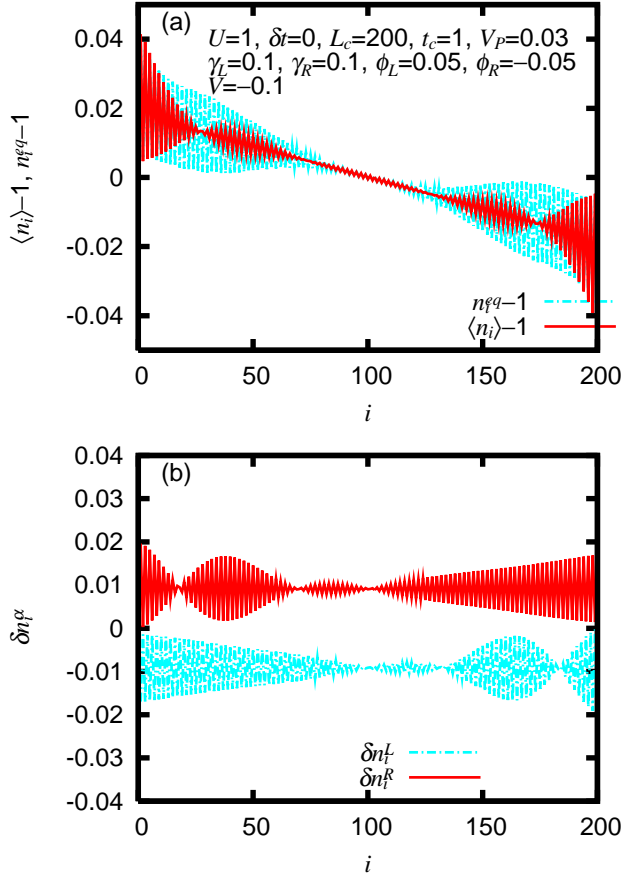}
\caption{(Color online) (a) Charge density, $ \langle n_i \rangle -1$ (red or gray), and ``equilibrium'' part, $ n^{\mathrm{eq}}_i -1$ (light blue or light gray), and (b) ``nonequilibrium'' part due to coupling with left electrode, $ \delta n^L_i $ ($<$0, light blue or light gray), and that with right electrode, $ \delta n^R_i $ ($>$0, red or gray), for Mott insulator ($ U $=1, $ \delta t $=0) with left-going bias ($ V $=$-$0.1). The other parameters are $ L_C $=200, $ t_c $=1, $ V_P $=0.03, $ \gamma_L $=$ \gamma_R $=0.1, and $ \phi_L $=$ -\phi_R $=0.05. 
\label{fig:Mott_negative_V}}
\end{figure}

Observing the charge distribution will help to understand the mechanism for making such a characteristic difference between Mott and band insulators. Figure~\ref{fig:Mott_negative_V}(a) shows the charge density, $ \langle n_i \rangle -1$=$ \sum_\sigma \langle n_{i\sigma} \rangle -1$, and its ``equilibrium'' part, $ n^{\mathrm{eq}}_i -1$=$ \sum_\sigma n^{\mathrm{eq}}_{i\sigma} -1$, for a Mott insulator with a left-going bias and barriers. 
Electrons accumulate near the left interface, while holes accumulate near the right interface. This is because the bands are bent near these interfaces (Fig.~\ref{fig:potential}). It should be noted that electrons and holes already accumulate for $ V $=0, which is responsible for the respective band bending at the interfaces and the resultant matching of the chemical potentials to reach equilibrium from the isolated case with $ \gamma_\alpha $=0. By giving a finite bias, a certain number of electrons come in and the same number of electrons (holes) go out (come in) in the steady state, so that the incoming electrons or holes cause no instability. If one takes a closer look, the charge density alternates between even and odd sites, although its amplitude is very small. This 2$ k_F $ oscillation is induced by the boundaries. 

Figure~\ref{fig:Mott_negative_V}(b) shows the ``nonequilibrium'' parts of the charge density, one of which is due to the coupling with the left electrode, $ \delta n^L_i $=$ \sum_\sigma \delta n^L_{i\sigma} $, and the other is due to that with the right electrode, $ \delta n^R_i $=$ \sum_\sigma \delta n^R_{i\sigma} $. Here, the voltage is negative so that electrons are left-going on average. Because of the coupling with the left electrode, electrons go out and $ \delta n^L_i <$0. Owing to the coupling with the right electrode, electrons come in and $ \delta n^R_i >$0. Thus, the present definition of $ \delta n^\alpha_i $ is useful for distinguishing between these flows. The property unique to Mott insulators is that  $ \delta n^\alpha_i $ is extended over the whole system in such a manner that $ \delta n^\alpha_i $ is almost constant except the 2$ k_F $ oscillation. Mott insulators disfavor further accumulation of charges relative to that in the case of $ V $=0, so that incoming electrons and holes are extended over the system. 

\begin{figure}[tb]
\includegraphics[height=12cm]{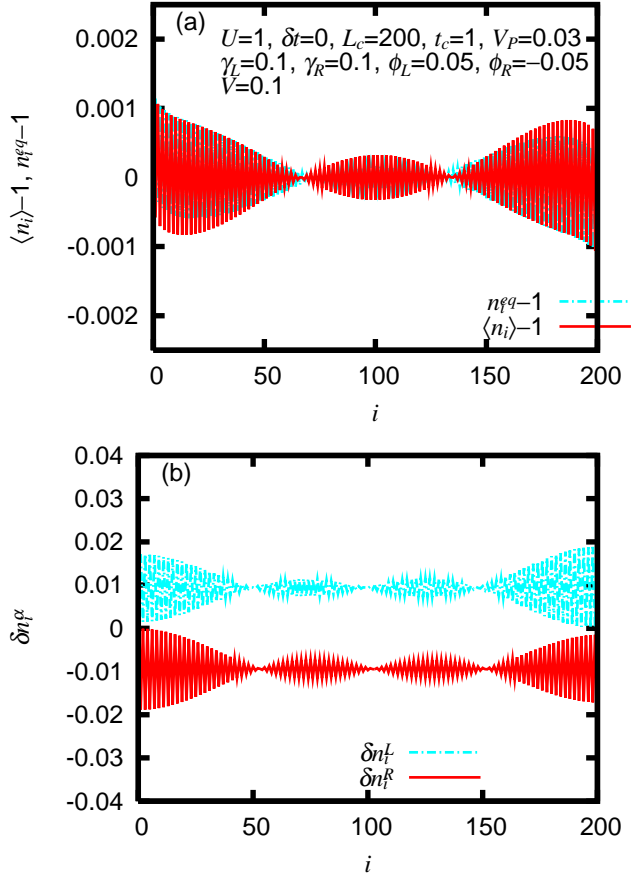}
\caption{(Color online) (a) Charge density, $ \langle n_i \rangle -1$ (red or gray), and ``equilibrium'' part, $ n^{\mathrm{eq}}_i -1$ (light blue or light gray), and (b) ``nonequilibrium'' part due to coupling with left electrode, $ \delta n^L_i $ ($>$0, light blue or light gray), and that with right electrode, $ \delta n^R_i $ ($<$0, red or gray), for Mott insulator ($ U $=1, $ \delta t $=0) with right-going bias ($ V $=0.1). The other parameters are the same as those in Fig.~\ref{fig:Mott_negative_V}. 
\label{fig:Mott_positive_V}}
\end{figure}

Figure~\ref{fig:Mott_positive_V} shows $ \langle n_i \rangle -1$, $ n^{\mathrm{eq}}_i -1$, $ \delta n^L_i $, and $ \delta n^R_i $ for a Mott insulator with a right-going bias. 
For this particular set of ($ \phi_L $, $ \phi_R $, $ V $)=(0.05, $-$0.05, 0.1), the potential $ \psi_i $ is constant everywhere [eq.~(\ref{eq:boundary})] and the density $ \langle n_i \rangle $ is almost unity (note the vertical scale) except for the 2$ k_F $ oscillation [eq.~(\ref{eq:poisson})]. Here, the voltage is positive so that electrons are right-going on average. Electrons come in from the left, $ \delta n^L_i >$0, and go out to the right, $ \delta n^R_i <$0. Both $ \delta n^L_i $ and $ \delta n^R_i $ are almost constant except for the small 2$ k_F $ oscillation. In Mott insulators, the averaged (i.e., 2$ k_F $-oscillation smoothed out) densities of the incoming electrons and holes show not only a small spatial modulation but also insensitivity to the sign of $ V $. This fact ensures the suppression of rectification because the current is given by the difference between the density of the incoming electrons at the exit and that of the outgoing electrons at the entrance [multiplied by the respective $ \gamma_\alpha $, eq.~(\ref{eq:current})]. 

\begin{figure}[tb]
\includegraphics[height=12cm]{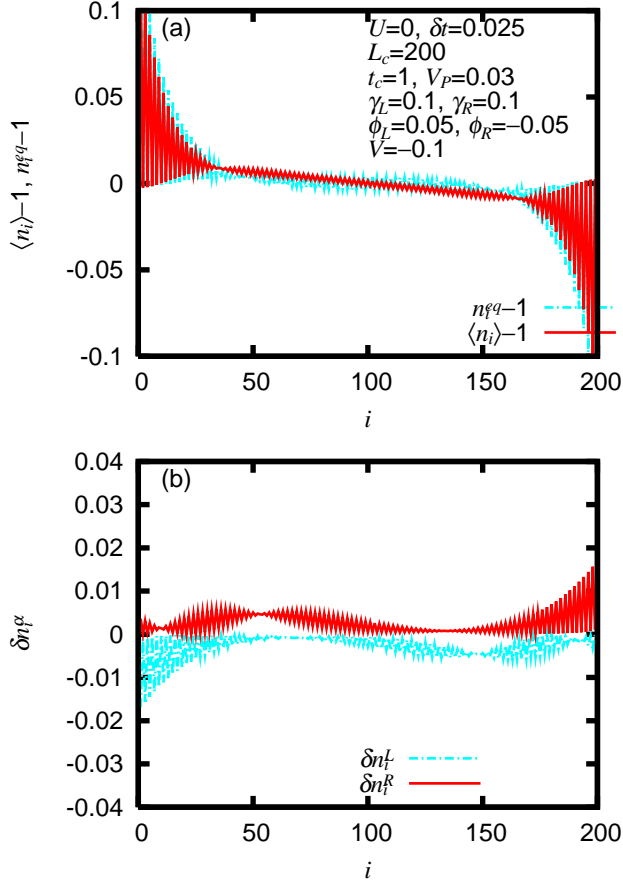}
\caption{(Color online) (a) Charge density, $ \langle n_i \rangle -1$ (red or gray), and ``equilibrium'' part, $ n^{\mathrm{eq}}_i -1$ (light blue or light gray), and (b) ``nonequilibrium'' part due to coupling with left electrode, $ \delta n^L_i $ ($<$0, light blue or light gray), and that with right electrode, $ \delta n^R_i $ ($>$0, red or gray), for band insulator ($ U $=0, $ \delta t $=0.025) with left-going bias ($ V $=$-$0.1). The other parameters are the same as those in Fig.~\ref{fig:Mott_negative_V}. 
\label{fig:band_negative_V}}
\end{figure}

The behavior of the metal--Mott-insulator interfaces is contrasted with that of the metal--band-insulator interfaces. Figure~\ref{fig:band_negative_V} shows $ \langle n_i \rangle -1$, $ n^{\mathrm{eq}}_i -1$, $ \delta n^L_i $, and $ \delta n^R_i $ for a band insulator with a left-going bias and barriers. 
It is clearly shown in $ n^{\mathrm{eq}}_i -1$ that electrons accumulate near the left interface, while holes accumulate near the right interface [Fig.~\ref{fig:band_negative_V}(a)]. They are much larger than the corresponding quantities at the metal--Mott-insulator interfaces. Thus, the accumulation is more sensitive to the band bending. 

The difference from the metal--Mott-insulator interfaces is conspicuous in $ \delta n^L_i $ and $ \delta n^R_i $ [Fig.~\ref{fig:band_negative_V}(b)]. Here, electrons are left-going on average. The quantity $ \delta n^L_i $ is thus negative, but its magnitude becomes almost zero at some point. On its left side, $ -\delta n^L_i $ is largest at the left interface and decays with increasing distance from the interface. This behavior originates from the outgoing electrons. On its right side, $ -\delta n^L_i $ shows a maximum away from the right interface. This behavior can be regarded as due to $ \psi_i $, which is higher on the right side. In other words, holes further accumulate on the right side at a negative $ V $. The spatial dependence of $ \delta n^R_i $ is obtained by the particle-hole transformation and the space-inversion operation of $ \delta n^L_i $. The quantity $ \delta n^R_i $ is positive and becomes almost zero at another point. On its right side, $ \delta n^R_i $ is largest at the right interface and decays with increasing distance from the interface. This behavior originates from the incoming electrons. On its left side, $ \delta n^R_i $ shows a maximum away from the left interface owing to further accumulation of electrons at lower $ \psi_i $'s. This further accumulation is realized by the absence of on-site repulsion. 

\begin{figure}[tb]
\includegraphics[height=12cm]{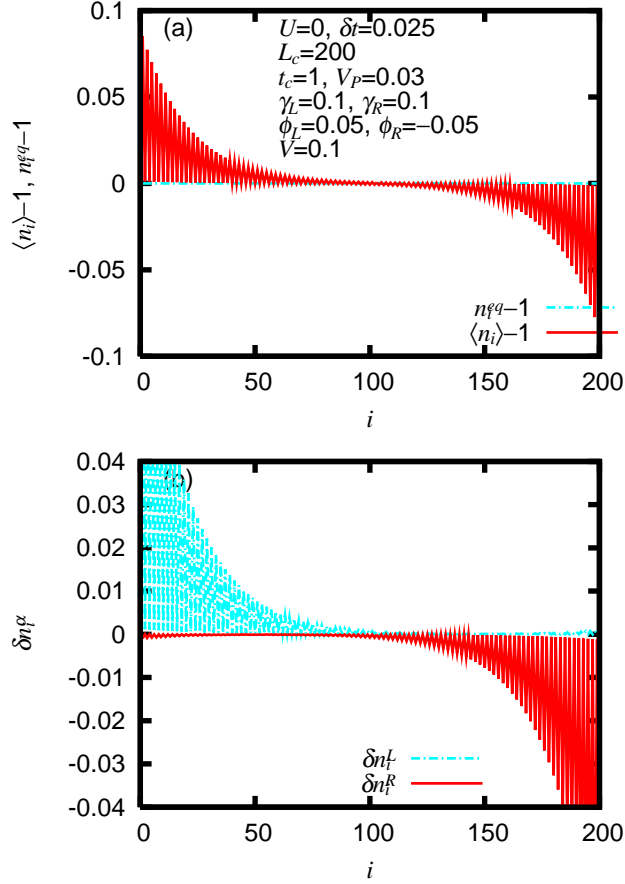}
\caption{(Color online) (a) Charge density, $ \langle n_i \rangle -1$ (red or gray), and ``equilibrium'' part, $ n^{\mathrm{eq}}_i -1$ (light blue or light gray), and (b) ``nonequilibrium'' part due to coupling with left electrode, $ \delta n^L_i $ ($>$0, light blue or light gray), and that with right electrode, $ \delta n^R_i $ ($<$0, red or gray), for band insulator ($ U $=0, $ \delta t $=0.025) with right-going bias ($ V $=0.1). The other parameters are the same as those in Fig.~\ref{fig:Mott_negative_V}. 
\label{fig:band_positive_V}}
\end{figure}

Figure~\ref{fig:band_positive_V} shows $ \langle n_i \rangle -1$, $ n^{\mathrm{eq}}_i -1$, $ \delta n^L_i $, and $ \delta n^R_i $ for a band insulator with a right-going bias. 
For this particular set of ($ \phi_L $, $ \phi_R $, $ V $), the potential $ \psi_i $ is constant. The quantity $ n^{\mathrm{eq}}_i $ is unity, but $ \langle n_i \rangle $ largely deviates from it. Because the density of electrons is much more sensitive to the bias-induced change in the potential distribution for band insulators, $ \langle n_i \rangle $ and $ n^{\mathrm{eq}}_i $ are now quite different. Here, electrons are right-going on average, so that $ \delta n^L_i >$0 and $ \delta n^R_i <$0. Because of the absence of barriers, a substantial number of electrons come in and go out. 

In order to see how the rectification is realized in a band insulator, we show its $ \delta n^L_i $ and $ \delta n^R_i $ in Fig.~\ref{fig:band_various_V} with parameters used in Fig.~\ref{fig:MW_IdnV}(b) and different biases, whose magnitudes are larger than the gap $ \Delta $=0.18 here.
For $ V < - \Delta $, the absolute value of $ J $, $ \mid J \mid $, is smaller than that for $ V > \Delta $ [Fig.~\ref{fig:MW_IdnV}(b)]. In this case, the density of the incoming electrons $ \delta n^R_i > 0 $ decays with increasing distance from the right interface, while the density of the outgoing electrons $ \delta n^L_i < 0 $ decays with increasing distance from the left interface. With increasing $ \mid V \mid $, both $ \mid \delta n^L_i \mid $ and $ \mid \delta n^R_i \mid $ slightly increase at each site $ i $, but they decay in quite similar manners [Fig.~\ref{fig:band_various_V}(a)]. 

For $ V > \Delta $, on the other hand, $ \mid J \mid $ increases more rapidly with $ \mid V \mid $ than for $ V < - \Delta $. Now, the current substantially flows through the system. The density of the outgoing electrons $ \delta n^R_i < 0 $ decays with increasing distance from the right interface. However, with increasing $ \mid V \mid $, a substantial number of electrons are shown to penetrate from the left interface into the system [Fig.~\ref{fig:band_various_V}(b)]. Both $ \mid \delta n^L_i \mid $ and $ \mid \delta n^R_i \mid $ finally decay toward the right and left interfaces, respectively, giving small $ \mid \delta n^L_{L_C} \mid $ and $ \mid \delta n^R_1 \mid $ values compared with $ \mid \delta n^L_1 \mid $ and $ \mid \delta n^R_{L_C} \mid $, respectively. However, $ \mid \delta n^L_{L_C} \mid $ for $ V > \Delta $ is much larger than $ \mid \delta n^L_{L_C} \mid $ and $ \mid \delta n^R_1 \mid $ for $ V < - \Delta $ because of the penetration of a larger number of electrons into the system. These behaviors of $ \delta n^L_i $ and $ \delta n^R_i $ in a band insulator are in contrast to those in a Mott insulator, where the averaged (i.e., 2$ k_F $-oscillation smoothed out) densities of the incoming electrons and holes show not only little spatial modulation but also insensitivity to the sign of $ V $. The $ V $ dependence of $ \delta n^L_i $ and that of $ \delta n^R_i $ in the Mott insulator closely follow that of $ \delta n^L_{L_C} $ and that of $ \delta n^R_1 $ in Fig.~\ref{fig:MW_IdnV}(a), respectively. 

\section{Summary and Conclusions \label{sec:summary}}

To elucidate the suppression of rectification at metal--Mott-insulator interfaces, which is obtained by numerical solutions to the time-dependent Schr\"odinger equation in our previous study, \cite{yonemitsu_prb07b} we employ nonequilibrium Green's functions in this paper. We consider one-dimensional half-filled electron systems and use the mean-field Hubbard model for a Mott insulator and the transfer-alternating tight-binding model for a band insulator. Metallic electrodes consisting of noninteracting electrons are attached to the system, and their effects are incorporated into self-energies within the wide-band limit. We take account of work-function differences, which are responsible for the formation of Schottky barriers, by adding to the model a scalar potential that satisfies the Poisson equation. 

In the mean-field approximation, the retarded and advanced Green's functions are obtained by solving the eigenvalue equation for a complex symmetric matrix, whose imaginary part comes from the self-energies due to the couplings with the electrodes. For the electron density to be determined self-consistently, we modify the frequently used formula in ref.~\citen{brandbyge_prb02} so as to maintain some symmetry in the particular case used here. The modification is useful for distinguishing between an inflow and an outflow in the ``nonequilibrium'' part of the density. The current is given by the difference between these flows, which are measured at appropriate sites and multiplied by couplings with electrodes. 

By plotting the spatial dependence of the density of incoming electrons and holes, we clarify the difference between metal--Mott-insulator and metal--band-insulator interfaces. In Mott insulators, the incoming electrons and holes are extended over the whole system in such a manner that $ \delta n^L_i $ and $ \delta n^R_i $ are almost constant over $ i $=1, $ \cdots $, $ L_C $. Electrons/holes do not accumulate at any place, so that the bias-induced change in the density of electrons is extended. Thus, charge transport becomes collective. 

\section*{Acknowledgment}

This work was supported by Grants-in-Aid for Scientific Research (C) (No. 19540381), for Scientific Research (B) (No. 20340101), and ``Grand Challenges in Next-Generation Integrated Nanoscience" from the Ministry of Education, Culture, Sports, Science and Technology.

\appendix
\section{Alternative Approach to Band Bending \label{sec:alternative}}

As is well known in electromagnetics, the Poisson equation is equivalent to the long-range Coulomb interaction, 
$ \sum_{j\neq i} (V/\mid r_j-r_i \mid) \left( n_j - 1 \right) \left( n_i - 1 \right) $, where $ r_i $ is the position vector for site $ i $. Then, we can define the Hartree potential as 
\begin{equation}
\psi'_i = \sum_{j(\neq i)} \frac{V}{\mid r_j-r_i \mid} 
\left( \langle n_j \rangle - 1 \right) 
\;,
\end{equation}
and add a linear term to it as 
\begin{equation}
\psi_i = \psi'_i + a i + b 
\;,
\end{equation}
where the constants $ a $ and $ b $ are determined so that the potential $ \psi $ satisfies the boundary condition [eq.~(\ref{eq:boundary})]. In this way, we can treat the band bending at interfaces and the effect of the long-range Coulomb interaction on the insulator (e.g., charge ordering if it is strong) on the same footing. This method can be extended to include the screening effect. In addition, we found that its numerical convergence is generally better than that of solving the Poisson equation. 

\section{Densities in the Atomic Limit \label{sec:atomic_limit}}

In the limit of $ t_c + (-1)^i \delta t  \rightarrow $0, i.e., in the atomic limit, one can simplify the expressions for the density. The matrix $ H_{\mathrm{HF}\sigma} $ is diagonal and its elements are given by $ e^\sigma_i \equiv \psi_i + U \langle n_{i\bar{\sigma}}-1/2 \rangle $, so that the eigenvalues $ \epsilon^\sigma_m - i \gamma^\sigma_m/2 $ are written as 
$ e^\sigma_{i_\alpha} - i \gamma_\alpha/2 $ for $ m $=$ i_\alpha $ and as 
$ e^\sigma_m $ otherwise. Then, the ``equilibrium'' and ``nonequilibrium'' parts of the density at $ i $=$ i_\alpha $ are then written as 
\begin{equation}
n^{\mathrm{eq}}_{i_\alpha\sigma} = 
\frac{1}{\pi} \tan^{-1} 
\frac{2(\mu_C-\epsilon^\sigma_{i_\alpha})}{\gamma_\alpha} + \frac12
\;, 
\end{equation}
and 
\begin{equation}
\delta n^\alpha_{i_\alpha\sigma} = 
\frac{1}{\pi} \left[ 
\tan^{-1} \frac{2(\mu_\alpha-\epsilon^\sigma_{i_\alpha})}{\gamma_\alpha} - 
\tan^{-1} \frac{2(\mu_C-\epsilon^\sigma_{i_\alpha})}{\gamma_\alpha}
\right] 
\;. 
\end{equation}
The total density is thus written as 
\begin{equation}
\langle n_{i_\alpha\sigma} \rangle = 
\frac{1}{\pi} \tan^{-1} 
\frac{2(\mu_\alpha-\epsilon^\sigma_{i_\alpha})}{\gamma_\alpha} + \frac12
\;, 
\end{equation}
which is a reasonable expression in the atomic limit. 

\bibliography{63588}

\end{document}